\documentclass[12pt,preprint]{aastex}

\usepackage{graphicx}
\usepackage{graphics}
\usepackage{natbib}

\newcommand{\be}{\begin{equation}}
\newcommand{\ee}{\end{equation}}

\newcommand{\ba}{\begin{eqnarray}}
\newcommand{\ea}{\end{eqnarray}}

\newcommand\eg{\textit{e.g.,}}

\newcommand\ie{\textit{i.e.}}

\newcommand{\Bf}{{magnetic field\,}}
\newcommand{\Bfs}{{magnetic fields\,}}
\newcommand{\Ef}{{electric  field\,}}
\newcommand{\Efs}{{electric fields\,}}

\newcommand{\ms}{magnetosphere}

\newcommand{\Fermi}{{\it Fermi}}
\begin{document}

\title{The gamma-ray spectrum of Geminga  and the  inverse Compton model of pulsar high energy  emission}
\author{Maxim Lyutikov\\
Department of Physics, Purdue University, \\
 525 Northwestern Avenue,
West Lafayette, IN
47907-2036 }

\begin{abstract}
We reanalyze the Fermi
  spectra of the Geminga and Vela  pulsars. We find  that the spectrum of Geminga  above the break is  exceptionally well approximated by  a simple power law  without the exponential cut-off, making  Geminga's spectrum  similar   to that of  Crab.  Vela's broadband gamma-ray spectrum is equally well fit with both the exponential cut-off and the double power law shapes. 
  
In  the broadband double power-law fits,  for a typical Fermi spectrum of a bright $\gamma$-ray pulsar,  most of the errors accumulate due to the  arbitrary parametrization of the spectral roll-off. In addition, a  power law with an  exponential cut-off  gives an acceptable fit for  the underlying  double power-law  spectrum for a very broad range of parameters, making  such  fitting procedures   insensitive to the underlying Fermi photon spectrum. 
 
 Our results have  important implications for the mechanism of pulsar high energy emission.
A number of observed properties of $\gamma$-ray pulsars, \ie,  the  broken power law spectra without  exponential cut-offs and stretching in case of Crab beyond the maximal curvature limit, spectral breaks close to or exceeding the maximal  breaks due to curvature emission, a Crab  patterns of relative intensities of the leading and trailing pulses repeated in the $X$-ray and $\gamma$-ray regions,  all point to the  inverse Compton origin of the high energy emission from majority of  pulsars. 
\end{abstract}

\section{VERITAS detection of Crab pulsar:  a case for  inverse Compton scattering origin of $\gamma$-ray emission}
The recent launch of the \Fermi\ Gamma-Ray Space Telescope and subsequent detection of a large number of pulsars \citep{2010ApJS..187..460A} revolutionized our picture of the non-thermal emission from pulsars in the gamma-ray band from 100\,MeV up to about 10\,GeV. At even higher energies, in the very-high energy (VHE) band, the detection of the Crab pulsar at 25\,GeV by the Magic Collaboration  \citep{2008Sci...322.1221A} and recently at 120\,GeV by the VERITAS Collaboration  \citep{VERITASPSRDetection} in the very-high energy (VHE)  band  allow to stringently constrain the very-high-energy emission mechanisms in the case of the Crab pulsar. 

\cite{2011arXiv1108.3824L} have argued that in case of Crab  the inverse Compton scattering is the main emission mechanism of the very high energy emission.  This was based on the detection of 
Crab pulsar by VERITAS collaboration above 150 GeV  \citep{VERITASPSRDetection} with non-exponential cut-off above the spectral break. The non-exponential cut-off  was later confirmed by the MAGIC collaboration \citep[][which also preferred IC model over the curvature emission]{2011ApJ...742...43A}.

The curvature emission in pulsars is limited to energies below 
\be
\epsilon_{br} = (3 \pi)^{7/4}  { \hbar \over ( c e) ^{3/4} } \eta^{3/4} \sqrt{\xi} \,  {  B_{NS}^{3/4} R_{NS}^{9/4}\over P^{7/4}} 
\label{1}
\ee
where $R_L$ is the light cylinder radius, $P$ is pulsar period of rotation, $\xi$ is a dimensionless scaling parameter $\xi=R_c/R_L$, $R_c$ is the radius of curvature of \Bf lines, $B = B_{NS} (R_{NS}/R)^3$, where $B_{NS}$ is the magnetic field on the surface of the neutrons star and $R_{NS}$ the starÕs surface  and $\eta = E/B \leq 1$ is the relative  strength of the accelerating \Ef, \citep{2011arXiv1108.3824L}. 

\begin{figure}[h!]
\includegraphics[width=0.99\linewidth]{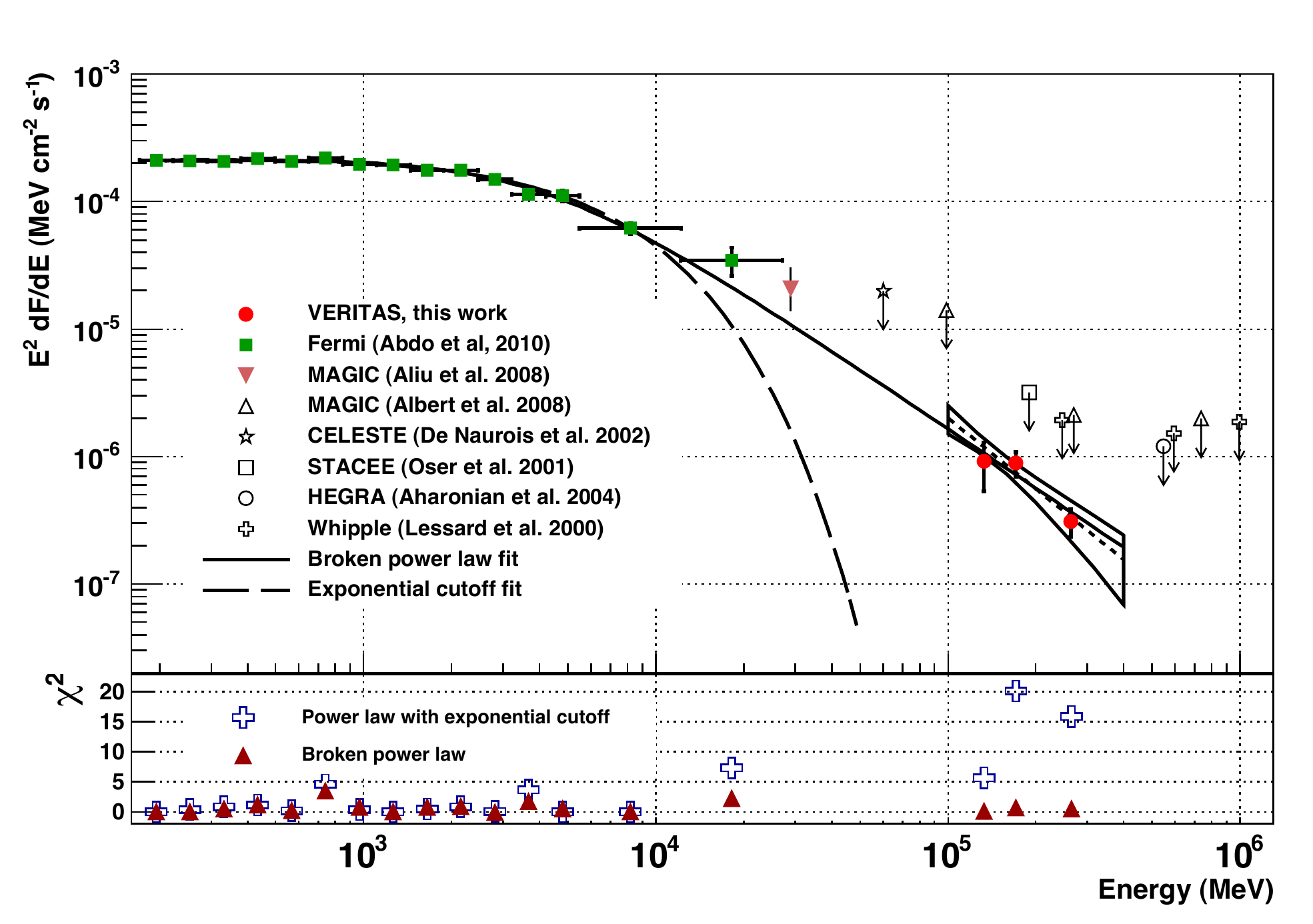}
\caption{High energy spectrum of Crab demonstrating the non-exponential spectral break inconsistent with curvature emission \citep[figure from][]{VERITASPSRDetection} . }
\label{crab-spectrum}
\end{figure}

If the $\gamma$-ray photons are due to the  curvature emission of a radiation reaction-limited population of leptons, the spectrum above the break must show an  exponential cut-off. 
The detection of the Crab pulsar by VERITAS collaboration   \citep{VERITASPSRDetection} clearly demonstrated the non-exponential cut-off above the spectral break, see Fig. \ref{crab-spectrum}. \cite{2011arXiv1108.3824L} have argued that this is inconsistent with the  curvature emission.

\section{Spectral breaks in  $\gamma$-ray pulsars}

The curvature and the IC model of the high energy emission offer different interpretations of the spectral breaks.
In case of the curvature emission, the spectral break corresponds to the maximal energy of the radiation reaction-limited acceleration. Above the break the spectrum is expected to have an exponential cut-off.
In the IC model the  spectral break corresponds to the break in particle distribution function, which is likely to be of the power law type both below and above the break. Thus, observations of the non-exponential cut-off above the break  favor the IC model.

Crab pulsar is a bright emitter in all spectral bands. Thus, there is a lot of low frequency target photons available for Compton scattering. What about other pulsars? In the present paper we argue that that there are evidence for a universal dominance of the inverse Compton scattering mechanism in $\gamma$-ray pulsars.

\begin{figure}[htb]
\includegraphics[width=0.99\linewidth]{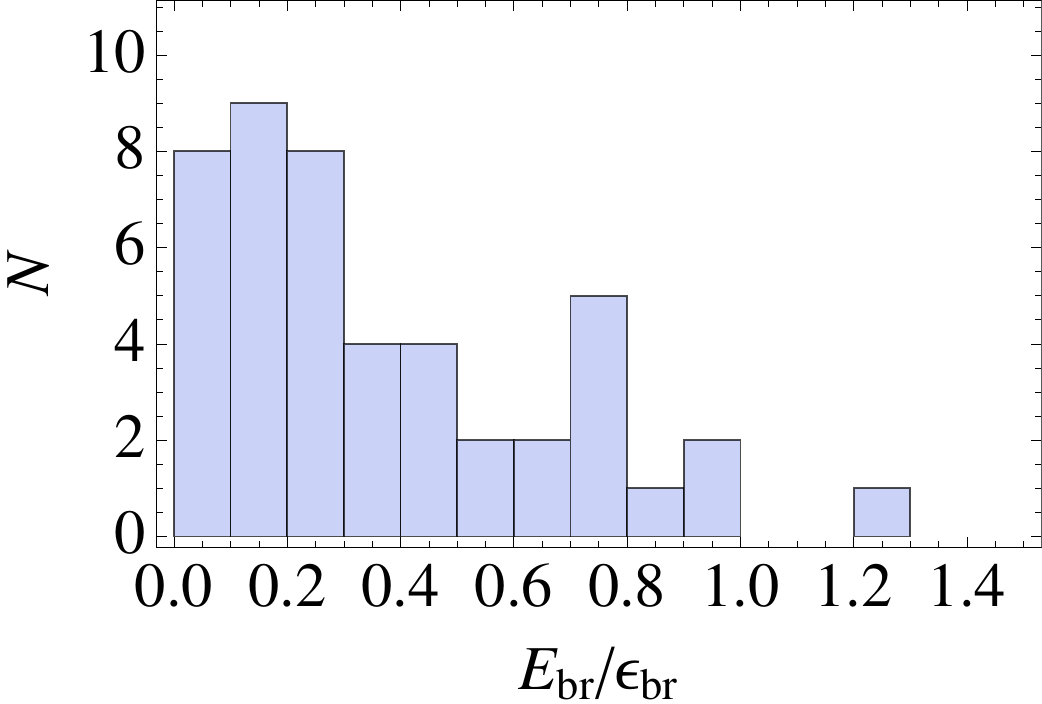}
\caption{Ratio of the  observed break energies $E_{br}$ for 46 pulsars to the maximum predicted for curvature radiation $\epsilon_{br}$, which is given by Eq.\ (\ref{1}) with $\eta =\xi=1$ \citep{2011arXiv1108.3824L}. }
\label{ratio}
\end{figure}

First, \cite{2011arXiv1108.3824L}  compared the observed spectral breaks of Fermi pulsars from the first Fermi catalogue \citep{2010ApJS..187..460A} with the predicted breaks due to curvature emission,  Eq.\ (\ref{1}) by calculating the ratio of the observed $E_{br}$ and predicted spectral break $\epsilon_{br}$, see Fig. \ref{ratio}.
A significant number of pulsars the ratio is close to one and for one pulsar, PSR  J1836 + 5925, the ratio is even larger than one. In order to explain the spectral break for these pulsars as a result of curvature radiation,  an accelerating \Efs\ should be close to or even larger than the \Bfs. 

What is more, the example of Crab demonstrates  that the spectral break may not be related to the maximal curvature photons.
The  presence of non-exponential breaks in pulsars others than Crab would be a clear indication that the break is not related to maximal energy of curvature emission by radiation-limited acceleration of leptons. Realistically, due to low photon counts, only the brightest $\gamma$-ray pulsars - Crab, Vela and Geminga - allow precise enough measurements of fluxes to distinguish  between different spectral shapes. In addition, typically, only phased average properties have enough photon counts to distinguish between the models. 

\section{Spectra of Geminga and Vela pulsars}

\subsection{Geminga:  a non-exponential cut-off}

\begin{figure}[h!]
\includegraphics[width=.99\linewidth]{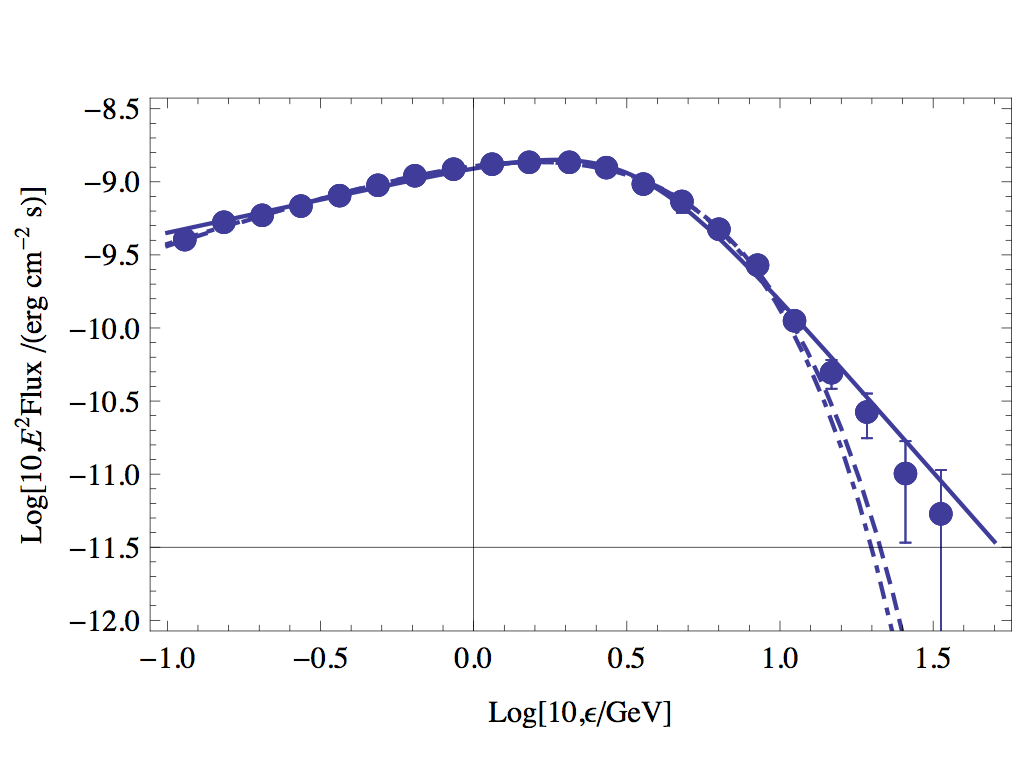}
\caption{Fits to Geminga spectrum by models described in Table \ref{Geminga-Fit}. Solid line   is  double power-law model  (a) from Table \ref{Geminga-Fit} (reduced $\chi^2 = 1.26$), dot dashed  is an exponential cut-off (reduced $\chi^2 = 1.13$) and dashed  is a softened  exponential cut-off (reduced $\chi^2 = 0.83$).  Since in all cases $\chi^2$ is of the order of unity, all three models are statistically acceptable.  (See also Fig. \ref{GemingaFit-err} for the error analysis of the   double power-law fit.)}
\label{Gem1}
\end{figure}

\begin{figure}[h!]
\includegraphics[width=.99\linewidth]{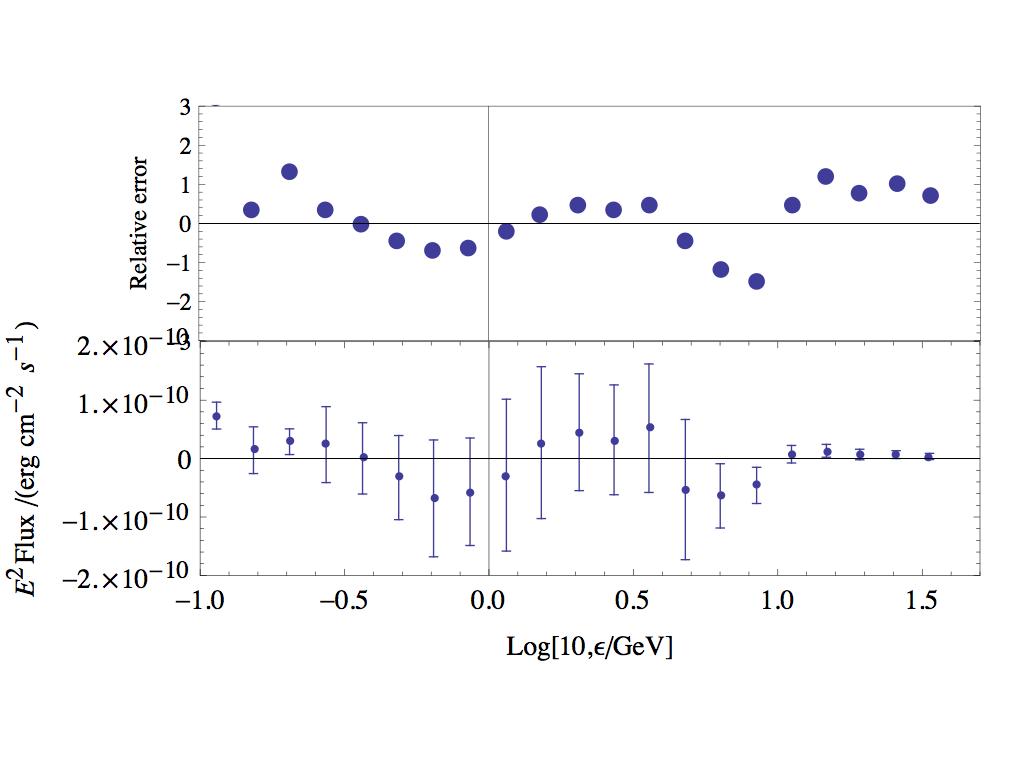}
\caption{ Errors for the broadband fit to Geminga data using double power law function. The lower  panel shows the residuals (on a linear scale) for the double power-law fit, while the upper panel shows relative errors that are used to calculate $\chi^2$. The errors are not random indicating the inadequacy  of the arbitrarily chosen spectral roll-off. Also,   most  of the $\chi^2$ is  accumulated  near the break energy due to the arbitrary parametrization of the spectral roll-off. Note that the highest energy data points actually have the smallest error bars. }
\label{GemingaFit-err}
\end{figure}

\begin{figure}[htb]
\includegraphics[width=0.99\linewidth]{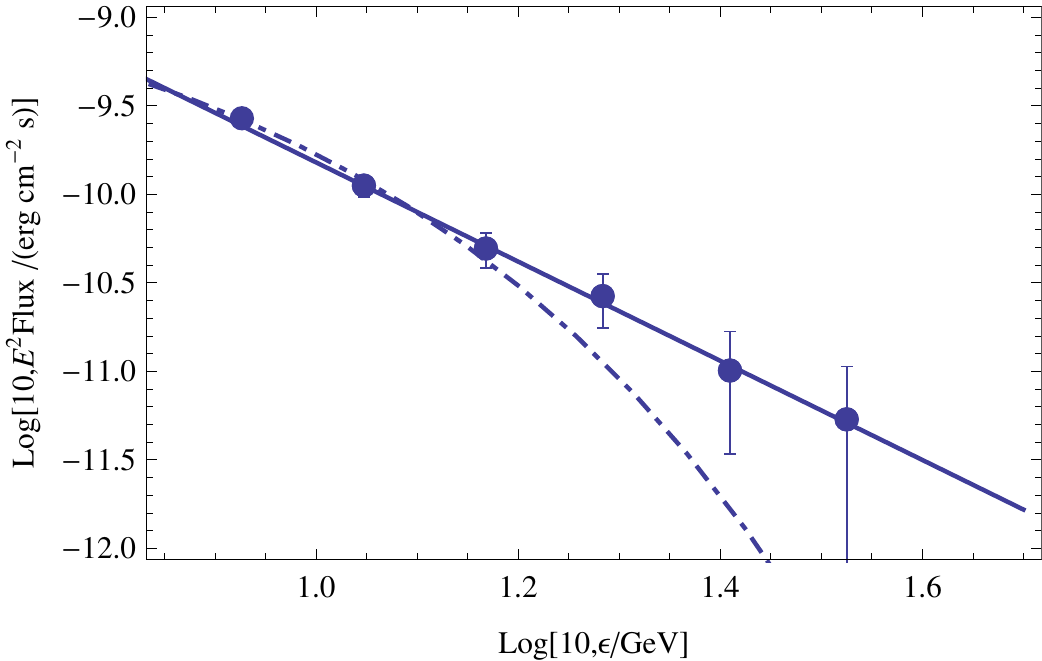}
\caption{Fits to the high energy tail of the  Geminga spectrum: power law (solid line, $\chi^2 =0.1$) and exponential cut-off (dashed line, $\chi^2 =2$).  }
\label{Gem2}
\end{figure}

Geminga is one of the brightest $\gamma$-ray pulsars. Its spectrum is, conventionally,  fit with a power-law plus exponential cut-off  \citep{2010ApJ...720..272A}.
We have performed an independent fit to Geminga spectrum, see Table \ref{Geminga-Fit}  and Figs. \ref{Gem1}-\ref{GemingaFit-err}-\ref{Gem2}. For the fits, the processed data (energy flux) provided by the Fermi collaboration were used, Table \ref{Gemin}. 

\begin{table}
\begin{tabular}{|c|c|c|}
\hline
E(GeV) & $E^2 $ \mbox{ Flux}  $(erg \,cm^{-2} s^{-1})$ & $dE^2 $ Flux $(erg \,cm^{-2} s^{-1})$ \\
\hline
0.114 & 4.02E-10 & 2.3E-11 \\
0.153 & 5.29E-10 & 4.0E-11 \\
0.204 & 5.90E-10 & 2.2E-11 \\
0.273 & 6.82E-10 & 6.5E-11 \\
0.365 & 8.04E-10 & 6.1E-11 \\
0.487 & 9.47E-10 & 7.2E-11 \\
0.643 & 1.10E-09 & 1.0E-10 \\
0.860 & 1.22E-09 & 9.2E-11 \\
1.149 & 1.32E-09 & 1.3E-10 \\
1.518 & 1.36E-09 & 1.3E-10 \\
2.053 & 1.36E-09 & 1.0E-10 \\
2.711 & 1.25E-09 & 9.4E-11 \\
3.580 & 9.65E-10 & 1.1E-10 \\
4.785 & 7.34E-10 & 1.2E-10\\
6.319 & 4.74E-10 & 5.5E-11 \\
8.446 & 2.69E-10 & 3.1E-11 \\
11.153 & 1.12E-10 & 1.5E-11 \\
14.728 & 4.94E-11  & 1.1E-11 \\
19.214 & 2.66E-11  & 9.0E-12 \\
25.681 & 1.01E-11  & 6.7E-12\\
33.505 & 5.35E-12 & 5.3E-12 \\
\hline
\end{tabular}
\label{Gemin}
\caption{Phase-averaged data for Geminga pulsar.}
\end{table}

First,  we performed  $\chi^2$ fits of the whole spectrum using a  particular type of  double power-law prescription (with 4 independent parameters, row (a) in Table \ref{Geminga-Fit},   the power-law plus exponential cut-off (with 3 independent parameters, row (b) in Table \ref{Geminga-Fit})  and a softened exponential cut-off (with 4 independent parameters, row (c) in Table \ref{Geminga-Fit}), see Table  \ref{Geminga-Fit} and Fig. \ref{Gem1}.   Weighted and unweighted reduced $\chi^2$  fits were not statistically distinguishable. (Overall normalization is the additional fit parameter to the ones listed in Table \ref{Geminga-Fit}.)

\begin{table}[h!]
$
 \begin{array}{|c|c|c|c|c|c|c|c|}
 \hline
\mbox{model}& \mbox{ Fit function} &  \alpha &  \beta  & \epsilon_{br}  &\mbox{ reduced, unweighted }   \chi^2  & b &   \mbox{dof} \\  \hline
a& \left( \left(  {\epsilon \over \epsilon_{br}} \right)^\alpha +  \left( {\epsilon \over \epsilon_{br}} \right)^{-\beta} \right)^{-1}  & 2.38 & 0.45 & 3.32& 1.26 & -  & 17 \\ \hline
b&  {\epsilon }  ^\beta e^{-{\epsilon \over \epsilon_{br}} }  & - & .70 & 2.35 & 1.13&  - &18\\ \hline
c&   {\epsilon }  ^\beta e^{- \left( {\epsilon \over \epsilon_{br}}\right)^b} & - & 0.75 & 1.98 & 0.83 &  0.91 &17  \\ \hline
\end{array}
$
\caption{Fits to the broad band spectrum of Geminga. The  values  for fit (c) can be compared with the best fit done by Fermi team, $ \beta=1.12,\,  \epsilon_{br} = 1.58, \, b=0.81$, \citep{2010ApJ...720..272A}.}
\label{Geminga-Fit}
\end{table}

Inspection of Table \ref{Geminga-Fit} and Figs. \ref{Gem1}-\ref{GemingaFit-err} tells,  first, that  since the reduced $\chi^2$ is close to unity all three models are statistically acceptable. (In Appendix \ref{append} we also demonstrate that for a typical Fermi spectrum of a bright $\gamma$-ray pulsar a  fit with an exponential cut-off can actually  give a satisfactory approximation to the underlying  power-law spectrum  for a very wide regime of fit parameters, \eg\  over two orders of magnitude in break energy.)

Second, 
the fit that looks  the best down to low flux levels, fit (a),  actually has largest reduced $\chi^2$ among the models.   The errors for the this fit  are  accumulated at intermediate energies due to {\it an arbitrarily chosen parametrization of the spectral roll-off.} Since we are mostly interested in the high energy scaling of the spectrum,  we can take only the highest energy points.
Having chosen the last 6 points we derive the fit given in Table \ref{Geminga-Fit-high}, see also Fig. \ref{Gem2}. 
\begin{table}[h!]
$
 \begin{array}{|c|c|c|c|}
 \hline
 \mbox{ Fit function} &  \alpha  &\mbox{ reduced }   \chi^2 & \mbox{dof}   \\  \hline
\epsilon ^\alpha & 3.04 & 0.1   &  4 \\ \hline
\end{array}
$
\caption{Properties of the fit of the high energy tail of Geminga.}
\label{Geminga-Fit-high}
\end{table}

Thus, the high energy part of the Geminga spectrum is exceptionally well fit with a power-law function, giving reduced $\chi^2= 0.1  $. This exceptional  small  value of $\chi^2$ is a bit surprising:   despite the lower overall count {\it the highest energy data points have the smallest error bars}, Table \ref{Gemin},  due to the very low background counts above $\sim 2 $ GeVs.  
 A small value of $\chi^2$ for the  high energy fit   may be due to the fact that the fit has only four degrees of freedom - in this case the random chance for all the points being very close to the model is not   small. This may also be an indication that the number of energy channels was chosen not in an optimal way, \eg\  too broad energy bins. (If only 5 highest energy points are used, the fit values are $ \alpha  = 2.8, \, \chi^2=0.03$ for 3 degrees of freedom.)

\subsubsection{Geminga to be seen at very high gamma-ray energies?}

The spectral index   above the break $dN/dE \propto E^{-p}$ with $p\approx 5$ for Geminga is  steeper than that of the Crab pulsar \citep[$p=3.8$ above the break][]{2011arXiv1110.4352M}.  Geminga is nearly two and  a half times  brighter than Crab at the peak of its spectral distribution of  $\sim 1 GeV$,  but its has lower break energy  and most importantly the steeper spectrum. The expected flux at $\sim 120 $ GeV, the low energy threshold for the VERITAS Cherenkov telescope, is then 
few $\sim 10^{-8} $ MeV \, cm$^{-2}$, s$^{-1}$. This is approximately ten times lower than the flux from the Crab pulsar.  Geminga, still, has a  chance to be detected,  since unlike Crab it has  no strong  background from the  PWN. (It is not clear whether the   excess TeV flux  from the general direction of Geminga \citep{2008PhRvL.101v1101A} is related to the pulsar \citep[see, though][]{2008A&A...485..527S}).

\subsubsection{Geminga: Phase-resolved fit}

Phase variation in the cut-off energy can affect the overall spectral fit, {\it but not at the highest energies}. For example, \cite{2010ApJ...720..272A} cite  the cut-off energy variations between approximately 1 and 3 GeV, while our high energy fit starts at 8.5 GeV, nearly three times higher, see Table \ref{Geminga-Fit-high} and Fig. \ref{Gem2}.  In addition, due to a low photon count in the phase-resolved SEDs, they  naturally miss the highest energy, lowest counts  tails.

\begin{figure}[htb]
\includegraphics[width=0.99\linewidth]{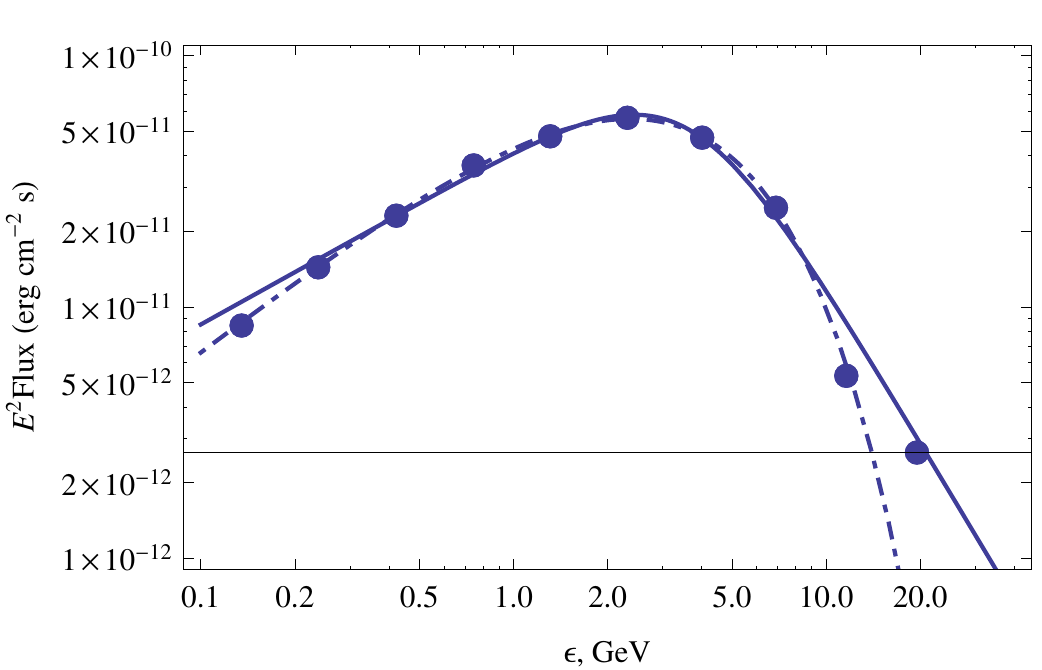}
\caption{Phase-resolved fit  to Geminga spectrum using double power law (solid line, reduced $\chi^2 = 0.46$)  and exponential cut-off (dot dashed line, reduced $\chi^2 = 0.23$). The fit corresponds to the phase $0.614 < \phi < 0.623$ of  \cite{2010ApJ...720..272A}.}
\label{Gem-resolved}
\end{figure}

To test the spectral shape in phased-resolved spectra we have performed spectral fitting at one particular phase,  $0.614 < \phi < 0.623$, corresponding, approximately to the second peak \cite{2010ApJ...720..272A}. Since phase resolved data were not available, we assumed the error on the flux to be  equal to $2 \times 10^{-12} {\rm erg\, cm^{-2} \,  s^{-1}} $,  see Table \ref{Gem-resolved-2} and Fig. \ref{Gem-resolved}.

\begin{table}[h!]
$
 \begin{array}{|c|c|c|c|c|c|c|c|}
 \hline
\mbox{model}& \mbox{ Fit function} &  \alpha &  \beta  & \epsilon_{br}  &\mbox{ reduced, unweighted }   \chi^2  & b &  \mbox{dof} \\  \hline
a& \left( \left(  {\epsilon \over \epsilon_{br}} \right)^\alpha +  \left( {\epsilon \over \epsilon_{br}} \right)^{-\beta} \right)^{-1}  & 2.08 & 0.69 & 3.61 & 0.46 & - & 6 \\ \hline
b&  {\epsilon }  ^\beta e^{-{\epsilon \over \epsilon_{br}} }  & - & 0.97 & 2.43 & 0.23&  -& 7 \\ \hline
\end{array}
$
\caption{Parameters for the phase-resolved fit of Geminga.}
\label{Gem-resolved-2}
\end{table}

Thus, both the double power-law and the exponential cut-off provide a statistically viable description of data for a particular  phase-resolved spectrum of Geminga.

\subsection{Vela: consistent with  the non-exponential tail}

\begin{figure}[htb]
\includegraphics[width=0.99\linewidth]{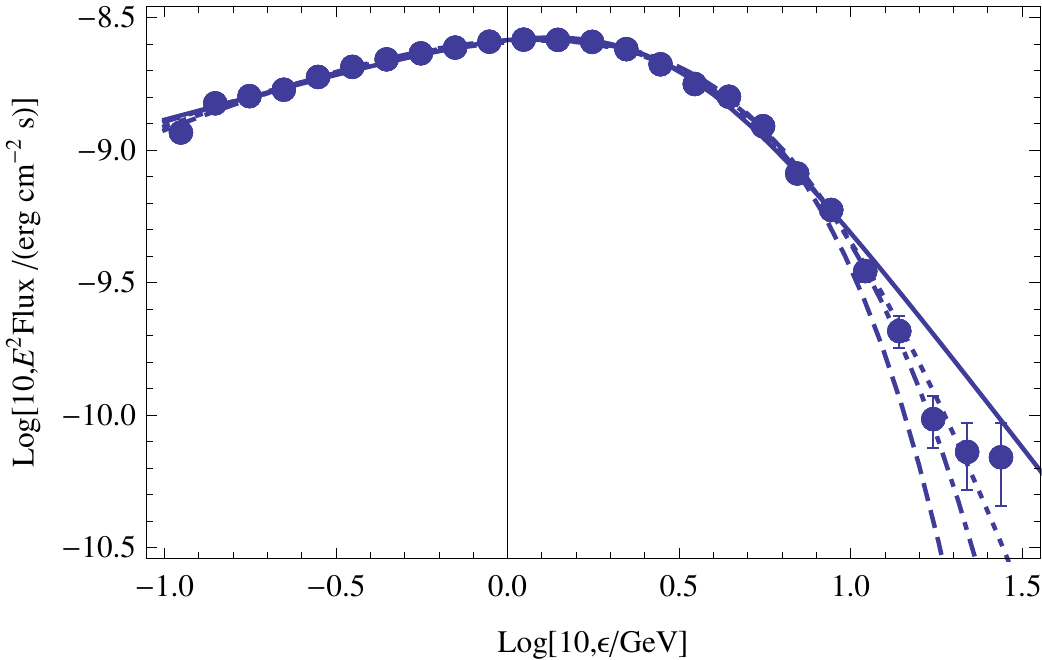}
\caption{Fits to Vela spectrum by models described in Table \ref{Vela-Fit}. Solid line: model (a), dotted line: model (b), dashed line: model (c), dot dashed line:  model (d) from Table \ref{Vela-Fit}.
Models (a) and (b) are double power-laws, (c) is an exponential cut-off and (d) is a softened  exponential cut-off. Models (a) and (c), as well as models (b) and (d) have approximately the same reduced $\chi^2$.}
\label{Vela-spectrum}
\end{figure}

Vela, the brightest $\gamma$-ray pulsar, offers tantalizing, but non-conclusive evidence of the non-exponential break, see Fig.  \ref{Vela-spectrum}.
 We performed an unweighted $\chi^2$ fit  with phase-averaged data provided by the Fermi collaboration, Table \ref{Vel}. We used various double power-law prescriptions (with 4 independent parameters), the power-law plus exponential cut-off (with 3 independent parameters)  and a softened exponential cut-off (with 4 independent parameters), see Table  \ref{Vela-Fit}.   
  Spectrum of Vela is equally well fit with a  softened exponential cut-off  and a double power-law spectrum.

\begin{table}
\begin{tabular}{|c|c|c|}
\hline
  E (GeV) &       E$^2$dN/dE  ($GeV\, cm^{-2} \, s^{-1}$) & $\Delta$  E$^2$ dN/dE  ($GeV\, cm^{-2} \, s^{-1}$)  \\
  \hline
     1.117101E-01 &  1.162739E-09 & 3.340227E-11 \\
    1.406211E-01 &  1.498706E-09 & 2.520782E-11 \\
    1.770114E-01 &  1.595459E-09 & 2.106914E-11 \\
    2.228146E-01 &  1.686333E-09 & 1.978347E-11 \\
    2.804637E-01 &  1.884574E-09 & 1.960196E-11 \\
    3.530192E-01 &  2.056793E-09 & 2.013993E-11 \\
   4.443313E-01 & 2.195488E-09 & 2.126256E-11 \\
    5.592426E-01 &  2.313047E-09 & 2.269755E-11 \\
    7.038429E-01 &  2.432161E-09 & 2.455590E-11 \\
    8.857895E-01 &  2.559430E-09 & 2.698647E-11 \\
    1.114708E+00 &  2.605548E-09 & 2.954981E-11 \\
    1.402695E+00 &  2.601159E-09 & 3.249238E-11 \\
    1.764951E+00 &  2.555662E-09 & 3.546979E-11 \\
    2.220567E+00 &  2.399446E-09 & 3.841962E-11 \\
   2.793512E+00 &  2.103038E-09 & 4.017982E-11 \\
    3.513868E+00 &  1.772092E-09 & 4.120424E-11 \\
    4.419371E+00 & 1.587057E-09 & 4.349206E-11 \\
    5.557327E+00 &  1.228406E-09 & 4.282005E-11 \\
    6.987010E+00 &  8.144106E-10 & 3.878978E-11 \\
    8.782625E+00 &  5.932086E-10 & 3.699830E-11 \\
    1.103701E+01 &  3.489809E-10 & 3.221406E-11 \\
    1.386622E+01 &  2.072424E-10 & 2.828281E-11 \\
    1.741518E+01 & 9.641950E-11  & 2.123657E-11 \\
    2.186479E+01 &  7.252995E-11  & 2.044876E-11 \\
    2.744061E+01 &  6.922606E-11  & 2.403077E-11 \\
   \hline
\end{tabular}
\label{Vel}
\caption{Phase-averaged data for Vela pulsar.}
\end{table}

\begin{table}[h!]
$
 \begin{array}{|c|c|c|c|c|c|c|c|}
 \hline
\mbox{model}& \mbox{ Fit function} &  \alpha &  \beta  & \epsilon_{br}  &\mbox{ reduced }   \chi^2  & b &  \mbox{dof} \\  \hline
a& \left( \left(  {\epsilon \over \epsilon_{br}} \right)^\alpha +  \left( {\epsilon \over \epsilon_{br}} \right)^{-\beta} \right)^{-1}  & 1.68& 0.34 & 2.91& 2.2& - &21  \\ \hline
b& \left( \left(  {\epsilon \over \epsilon_{br}} \right) +\left( {\epsilon \over \epsilon_{br}} \right)^{-\beta} \right)^{-\alpha}  & 4.11 & 0.11& 8.8 & 1.3& - & 21\\ \hline
c&  {\epsilon }  ^\beta e^{-{\epsilon \over \epsilon_{br}} }  & - & .42 & 3.1 & 2.0&  -& 22 \\ \hline
d&   {\epsilon }  ^\beta e^{- \left( {\epsilon \over \epsilon_{br}}\right)^b} & - & 0.59 & 1.58 &1.4 &  0.73& 21  \\ \hline
\end{array}
$
\caption{Parameters of the spectral fit for Vela pulsar}
\label{Vela-Fit}
\end{table}

In case of Vela, a particular choice of the parametrization of the double power-law spectrum illustrates 
that an arbitrary choice of the roll-off between the two spectral power-law components has an important effect on the fit:  while the more conventional double power-law  parametrization in line (a), Table \ref{Vela-Fit},  gives unacceptable fit,  double power-law spectrum parametrization in line (b) in Table \ref{Vela-Fit} gives an acceptable fit.

\section{Optical-$X$-ray-$\gamma$-ray correlation}

\begin{figure}[htb]
\includegraphics[width=0.99\linewidth]{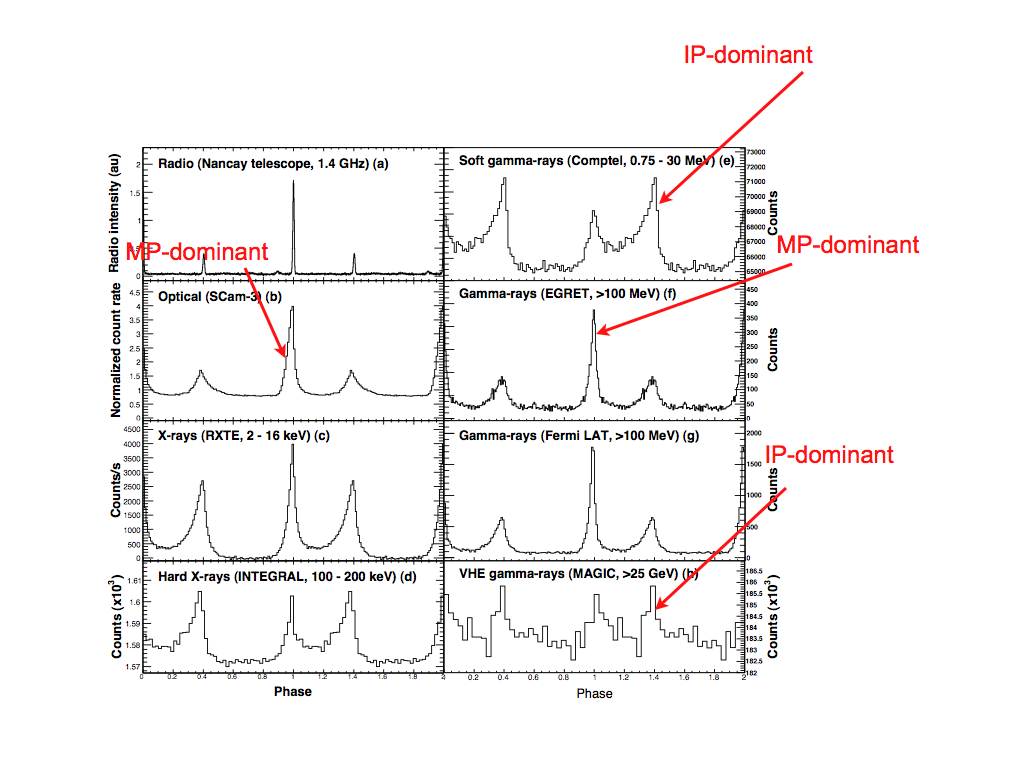}
\caption{Evolution of the Crab profile with energy \citep{2010ApJ...708.1254A}. Note that that the lower-energy evolution of the increasing interpulse to main pulse ratio is mirrored in the $\gamma$-rays. Such behavior is expected in synchrotron-self-Compton model \citep{2011arXiv1108.3824L}. }
\label{Crab-profile}
\end{figure}

Within the framework of the SSC model the power emitted by IC  is related to the power of the seed photons. Photons of different energies that are emitted by the same particles should in principle produce similar pulse profiles. In our model one expects, therefore, that the pulse profiles in X-ray and in gamma-rays are similar because the secondary plasma emits synchrotron radiation in X-rays and IC scatters UV photons into the VHE band. And indeed, the ratio of the amplitudes of the two pulses in the pulse profile of the Crab pulsar changes consistently in the X-rays / soft gamma-ray band and in the high energy gamma-ray band, see Fig. \ref{Crab-profile}. In X-rays the main pulse dominates over the inter pulse. The ratio changes towards higher energies and reverses in the soft gamma-ray band at about 1\,MeV. Mirroring the evolution at lower energy, the main pulse dominates at 100\,MeV, while at 120\,GeV the interpulse clearly again  dominates over the main pulse \citep{2010ApJS..187..460A,VERITASPSRDetection}.

\begin{figure}[h!]
\includegraphics[width=0.99\linewidth]{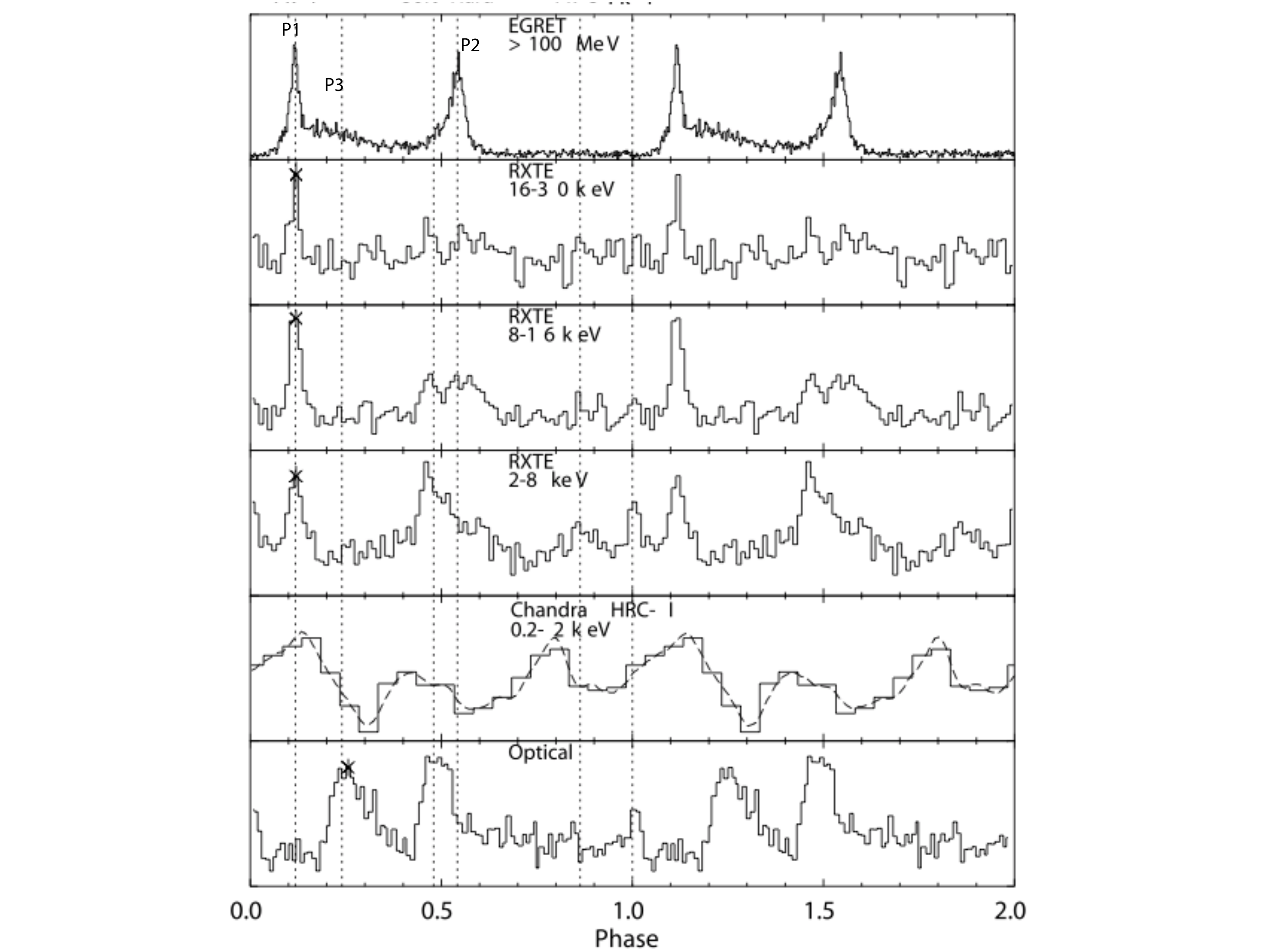}
\caption{Vela profile in broad energy bands \citep{2010ApJ...713..154A}. Chandra profile in the soft $X$-rays is dominated by thermal emission and is not of interest here. Optical and RXTE signals are mostly non-thermal. Note that all three $\gamma$-ray peaks (P1, P2 and P3 in the top panel) have lower frequency non-thermal  analogues: P1 in the $X$-rays, P3 in optical and P2 in both bands. This is expected in the IC model, though the spectra and  relative intensities depend on the details of the photon and particle spectra. }
\label{Vela-profile}
\end{figure}

Vela also offers tantalizing evidence for the correlation between the optical/$X$-ray  and $\gamma$-ray bands, Fig. \ref{Vela-profile}: all three $\gamma$-ray peaks (P1, P2 and P3 in the top panel) have lower frequency analogues: P1 in the $X$-rays, P2 in optical and P3 in both bands. This is expected in the IC model, though relative intensities and spectra of the components depend on the details of the photon and particle distribution.

Crab pulsar is somewhat exceptional among the brightest  $X$-ray-$\gamma$-ray pulsars since it has no indication of the thermal component. The $X$-ray emission of Vela and Geminga are strongly dominated by the thermal components coming form the surface \citep{1993ApJ...415..286H,2001ApJ...552L.129P,2007ApJ...669..570M}. In Vela, the non-thermal component in $X$-rays is barely detected  at some orbital phases, roughly aligned with the $\gamma$-ray peaks, especially with P2, see Fig. 9 of  \cite{2007ApJ...669..570M}. The relative weakness of the non-thermal $X$-ray components does not allow a measurement of the evolution of the P1/P2 ratio with energy, like in Crab.
Overall, both  Vela and Geminga  spectra show a complicated phase- and energy-dependent mix of thermal and non-thermal components \citep[\eg][]{2005ApJ...625..307K,2005ApJ...627..383R}. In both cases the optical emission, the main target for IC scattering, is highly non-thermal.

\section{Discussion}

In this paper we demonstrate that, first, broadband pulsar high energy spectra are generally consistent with the  double power law shapes. This results is based on three brightest $\gamma$-ray pulsars. In Geminga the high energy part of the SED is exceptionally well fit with a power law, while the spectrum of Vela is inconclusive. 
Importantly, in Geminga most errors in the broadband    double power law fit are accumulated due to  an {\it  arbitrary parametrization of the spectral roll-off between the two asymptotic power   laws}. Thus, the standard  $\chi^2$ fit  underestimates the goodness of fit in this particular case. In addition, the low background at high energies, above 2 GeV, make Fermi data especially sensitive (smallest error bars) to the shape of the high energy spectrum.

If pulsar spectra  generally  show non-exponential cut-offs, a fact well established in the Crab pulsar by VERITAS  \citep{VERITASPSRDetection} and in Geminga in the present paper, this has important implications for models of pulsar $\gamma$-ray  emission. This implies the  importance of the inverse Compton (IC) scattering for the production of $\gamma$-ray photons  \citep{2011arXiv1108.3824L}  and would   signify  a paradigm shift  in the study of pulsar high-energy emission.

  \cite{2011arXiv1108.3824L}  argued that in Crab and  in a number of other pulsars the inverse Compton scattering is the main emission mechanism of the very high energy emission.  The authors  tried to give the most general arguments in favor of IC,  mostly independent of the numerous possible particular details of a full radiative model. Overall,  \cite{2011arXiv1108.3824L} adopted the \cite{1986ApJ...300..522C} paradigm of pulsar high energy emission, but with inverse Compton scattering playing a more important role that previously assumed. Recall, that  \cite{1986ApJ...300..522C} argued that IC scattering is not dominant in Crab, but may be important for Vela. VERITAS and MAGIC results on Crab indicate that IC is the dominant mechanism in Crab.

  In the model  of \cite{2011arXiv1108.3824L}  the inverse Compton scattering occurring   in the Klein-Nishina regime by the secondary particles results in a picture that, overall, is  consistent  with observations without any fine-tuning. 
The key features of our model are (i) A population of primaries that is accelerated in a modest  electric field, which is a fraction $\eta$ of the magnetic field strength near the light cylinder with  a typical value of $\eta$ is  $10^{-2}$. The suppression of the scattering cross-section in the Klein-Nishina regime (and the corresponding  lower radiation loss rate of electrons)  allows primary  leptons to be accelerated to very high energies with hard spectra.  (ii) The gain in energy of the primaries in the electric field is balanced by similar curvature radiation and IC losses (radiation reaction limited); (iii)   The secondary plasma is less energetic, but more dense and has approximately the same energy content as the primary beam. The secondary plasma is responsible for the soft UV-$X$-ray emission via synchrotron/cyclotron emission and the high energy
$\gamma$-ray emission that extends to hundreds of GeV via the inverse Compton process. The IC emission from the primary beam extends well into the TeV regime but will be difficult to detected due to the low predicted fluxes.

The spectra of all three brightest $\gamma$-ray pulsars are consistent with the broken power-law distribution (and, thus, with the inverse Compton origin of the $\gamma$-ray emission). What is more, Crab and Geminga are inconsistent with exponential cut-off (and, thus, with the curvature origin of the $\gamma$-ray emission). Thus, inverse Compton scattering may be the  dominant  radiation mechanism in all pulsars, though a curvature contribution in some pulsars cannot be excluded. 

In the analysis we use the processed data points provided by the Fermi collaboration as well as data from the published papers. These data points are
actually somewhat dependent on the model used to derive them, since
the model of the spectrum goes into untangling the data from the
instrument response. This introduces some uncertainty, but since we are looking at the brightest pulsars we expect that  a particular model used to estimate
the flux in individual spectral bins is not going to make too much of a 
difference. We would like to encourage the community  to test our results using  the raw count rates data.

It is not clear what parameters determine the  dominance of the IC scattering   among  Vela, Crab and Geminga. Calculations of  \cite{1986ApJ...300..522C} suggested that  IC scattering may be important for Vela, but our results suggest that Vela is the weakest case for the IC among the three. 
The case for dominance of IC scattering in Crab was, partially, expected due to high soft photon luminosity \citep[\eg][]{2007ApJ...662.1173H}. 
What is more, the Crab emission is completely non-thermal and originates in the same region of the \ms\ as the  $\gamma$-ray emission, while in Vela and Geminga a large contribution to $X$-ray emission comes from the thermal emission from the surface. 

In addition, the strong energy dependence of IC scattering in the KN regime favors the  lower energy optical-UV photons.
In this view, 
  the example of non-exponential cut-off in Geminga, and the implied importance of the IC scattering, are quite surprising given  the low observed optical-$X$-ray  luminosity. This poses challenges for radiative modeling. It is likely to be related to the fact that  inverse Compton scattering
is highly  dependent on the details of both the soft photon distribution and the particle's  distribution (\eg\ their anisotropy).

Finally, let us recapitulate the main arguments in favor of inverse Compton scattering being the dominant  pulsar radiation mechanism at GeV energies: 
\begin{itemize}
\item The spectrum of Crab pulsar extends beyond the upper limit for curvature emission
\item The beak energy in  many pulsars approaches and even exceeds  the upper limit for curvature emission
\item In the three  brightest pulsars, Crab, Geminga and possibly Vela the spectrum above the break is not exponentially suppressed, implying that the break is not due to curvature remission of leptons in the radiation reaction-limited regime
\item In Crab, the energy dependence of the relative intensity of  emission peaks at $\gamma$-ray  energies mirrors the lower energy energy dependence, as expected in the Inverse Compton model.
\end{itemize}

I would like to thank Oleg Kargaltsev, Matthew Lister, Rafael Lang, George Pavlov, Scott Ransom and Mallory Roberts   for  help with the data analysis, Roger Blandford, John Finley, Andrew McCann, Roger Romani, Nepomuk Otte, Mark Thieling  and David Thompson for  comments on the manuscript and  Fermi Large Area Telescope Collaboration  for providing Vela \& Geminga  phase-averaged data. 

\bibliographystyle{apj}
  \bibliography{/Users/maxim/Home/Research/BibTex} 

\appendix
\label{append}

\section{On the exponential fit to Fermi-like   data with power-law distribution}

Often, Fermi pulsar data are fit with a softened exponential function, $ \propto \epsilon^\alpha e^{- ( \epsilon/ \epsilon_{br})^b}$, which is supposed to imitate variations in the exponential cut-off energy.  In this appendix we address a question: if the data come from double power-law distribution, how well will they be represented by the softened exponential fit? As an example, we fit a function $f =2 /(x +1/x)$ sampled at equally spaced logarithmic intervals from $0.1$ to $40$ with a typical error of $0.01$ and random scatter $0.01$. This  choice of the parameters  simulates a typical
bright pulsar spectrum sampled in unites of GeV and maximum flux of the order of $\sim 100$ of the statistical error. 

\begin{figure}[htb]
\includegraphics[width=0.99\linewidth]{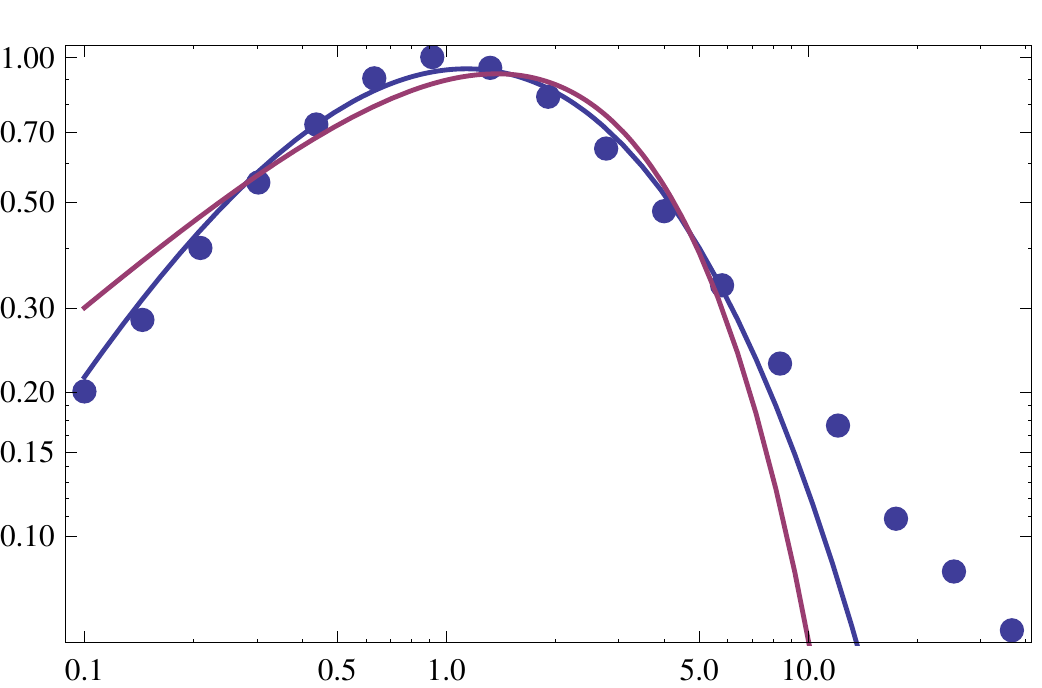}
\caption{Numerical fit to the function $f =2 /(x +1/x)$ using a  softened exponential function, $ \propto \epsilon^\alpha e^{- ( \epsilon/ \epsilon_{br})^b}$. Two  curves  correspond to extreme values of the parameters ($\epsilon_{br}=0.01,\, \alpha = 1.87,\, b=0.35$ and  $\epsilon_{br}=2,\, \alpha = 1.68,\, b=0.98$), but both give acceptable fits (reduced $\chi^2 = 0.33 $ and $1.01$ correspondingly.) }
\label{exp-fit}
\end{figure}

 We find that the fit function provides an acceptable fit (reduced $\chi^2 \leq 1$) for  an exceptionally wide range of the break energies $0.01 \leq \epsilon_{br} \leq 2$, see Fig. \ref{exp-fit}. 
 For less bright pulsars, where due to low photon statistics  the spectrum can be measured up to a  lower maximal energy, the reduced $\chi^2 $ remains smaller than unity even for larger values of the fitted $ \epsilon_{br} $.

Thus, if a typical spectrum of a bright Fermi pulsar  is a broken power-law, 
  the spectral fit with a softened exponential function, $ \propto \epsilon^\alpha e^{- ( \epsilon/ \epsilon_{br})^b}$,  gives acceptable fits for a very  wide range of parameters and, thus, does not provide a sufficiently sensitive probe of the underlying spectrum. 

\end{document}